\documentclass{amsart}
\usepackage{graphicx,hyperref}
\usepackage{verbatim}
\usepackage[english]{babel}

\usepackage[margin=2cm]{geometry}
\usepackage{lipsum}
\usepackage{amsmath}

\begin{document}

\title{The lunar cycle's influence on sex determination at conception in humans}

\author{Derek Onken}
\address{Derek Onken: Emory University, Department of Mathematics \& Computer Science, Atlanta, GA, 30322, USA}
\author[2]{Eric Marty}
\address{Eric Marty: University of Georgia, School of Arts, Athens GA, 30602, USA}
\author[3]{Roberto Palomares}
\address{Roberto Palomares: University of Georgia, Veterinary Medicine, Athens GA, 30602, USA}
\author[4]{Rui Xie}
\address{Rui Xie: University of Georgia, Department of Statistics, Athens GA, 30602, USA}
\author[5]{Leyao Zhang}
\address{Leyao Zhang: University of Georgia, Department of Statistics, Athens GA, 30602, USA}
\author[6]{Jonathan Arnold}
\address{Jonathan Arnold: University of Georgia, Department of Genetics, Athens GA, 30602, USA}
\author{Juan B. Gutierrez}
\address{Juan B. Gutierrez: \href{mailto:jgutierr@uga.edu}{jgutierr@uga.edu} University of Georgia, Department of Mathematics and Institute of Bioinformatics, Athens, 30602, USA}


\begin{abstract}
         The lunar cycle has long been suspected to influence biological phenomena. Folklore alludes to such a relationship, but previous scientific analyses have failed to find significant associations. 
         It has been shown that lunar cycles indeed have effects on animals; significant associations between human circadian rhythms and lunar cycles have also been reported. 
         We set out to determine whether a significant statistical correlation exists between the lunar phase and sex determination during conception. 
          We found that significant associations (\textit{p}-value $< 5 \times 10^{-5}$) exist between the average sex ratio (male:female) and the lunar month. 
         The likelihood of conception of a male is at its highest point five days after the full moon, whereas the highest likelihood of female conception occurs nineteen days after the full moon. Furthermore, we found that the strength of this influence is correlated with the amount of solar radiation (which is proportional to moonlight). 
         Our results suggest that sex determination may be influenced by the moon cycle, which suggests the possibility of lunar influence on other biological phenomena.  
        We suggest for future research the exploration of similar effects in other phenomena involving humans and other species.
        
        
     
         
         \end{abstract}

\flushbottom
\maketitle
%
%
\thispagestyle{empty}

\section*{Introduction} \label{sec:intro}

The lunar cycle has long been suspected to influence biological phenomena. Early myths associate the moon with growth efficiency of crops, animal reproduction and behavior, and human fertility. The ancient Greeks worshiped Artemis as the goddess of the moon, childbirth, and fertility \cite{littleton2002}. Aristotle, credited with one of the earliest references to the influence of the moon phase on reproduction (Parts of Animals, 680a:30), noted that the ``ova,'' or reproductive organs of the sea urchin, are much larger in the nights of the full moon, although modern research observes this only in the Red Sea \cite{fox1924lunar}. Additionally, Ptolemy mentions that the most copulation of herds occurs during full moon (Tetrabiblos Book I Chapter III). 

The Moon takes roughly 27.32 solar days to circle the Earth in what is called the Moon's \textit{sidereal cycle}. However, simultaneously, the Earth orbits the Sun and as a consequence the full cycle of the moon as observed from Earth exceeds the sidereal cycle in length\cite{rotton1985}; this \textit{synodic cycle} is currently estimated at 29.53 days. A \textit{metonic cycle} occurs when the solar year and the synodic cycle share a multiple; 19 solar years roughly (fewer than 2 hours off) equates to 235 synodic months, a calculation the Babylonians made with limited understanding of the planetary physics \cite{gutzwiller1998}. 
The phases of the Moon, as we know them, correspond to the synodic cycle; the ends and beginnings of each lunar month are dictated by a visible full moon, and each month can be easily divided into quarters or lunar phases \cite{rotton1985}.

It has been shown that lunar cycles indeed have effects on animals.  Honeybees peak in bodyweight during a new moon, and their triacyloglycerols and steroids follow a 29.5-day cycle \cite{zimecki2006}. Fish experience similar cyclicality. Periodogram analysis of the sperm cycle of the mummichog (\textit{Fundulus heteroclitus}) reveals a semilunar cycle of 15 days \cite{cochran1988}. Plasma melatonin levels of the golden rabbitfish (\textit{Siganus guttatus}) fluctuate with the moonlight intensity at night, peaking during new moon; furthermore, these fish do not spawn when light intensity is held constant \cite{zimecki2006}. Thyroxine levels of salmonid fish surge in the new moon in preparation for seaward migration \cite{grau1981}. There is evidence that this type of phenomenon is influenced by moonlight intensity rather than by gravitational effects; moreover, it has been well documented that the reproductive activities of many fish are influenced by moonlight \cite{takemura2004,takemura2010}.  Among vertebrates, a physiological response to moonlight is not exclusive to fish; it has been observed that lunar cycles and amount of light affect the nuclear size of the pinealocytes of Wistar rats \cite{martinez2002}. Zimecki (2006) \cite{zimecki2006} reports that Nazca boobies (\textit{Sula granti}) experience lowest concentrations of melatonin during the full moon. Navara and Nelson (2007) \cite{navara2007} describe how light encourages the pineal gland to suppress melatonin; furthermore, it is known that the presence of melatonin decreases estrogen secretion in many mammals. An investigation designed to test whether mare fertility varies with lunar cycles using two thoroughbred stud farm records over nine consecutive years revealed that the probability of conception was higher two days after full moon \cite{kollerstrom2000}.  Additionally, increased fertility of artificial insemination of dual purpose bovines has been reported during the full moon \cite{Perea2005}.

Associations between human circadian rhythms and lunar cycles have had mixed reviews \cite{foster2008}.  Ghiandoni \textit{et al}. (1998) \cite{ghiandoni1998} observed more deliveries on the first and second day after a full moon, especially for multiparous births, with a small sample of births in a hospital in Fano, Italy. This result was confirmed by Charpentier and Causuer (2009) \cite{charpentier2009} using 25.4 million births in France between 1968 and 2005. However, Strolego \textit{et al.} (1991) \cite{strolego1991} depart from this finding based on the study of births in Mozambique; Periti and Biagiotti (1994) \cite{periti1994} also found no association between lunar phase and delivery rates. Human fertility in New York City was highest in the  quarter of the lunar month immediately following the full moon \cite{zimecki2006}. Additionally, Cutler \textit{et al}. (1987) \cite{cutler1987} noticed a correlation of onset of menses in the same phase. Humans, among numerous other species beyond the Nazca booby, experience suppression of melatonin levels when exposed to extended periods of light \cite{navara2007}.  Specifically, ``[h]uman exposure to a low-level incandescent bulb at night requires only 39 min to suppress melatonin levels to 50\%''  \cite{navara2007}.
Das \textit{et al}. (2005) \cite{Das2005} found that the lunar phase had no effect on \textit{in vitro} fertilization.  Other skeptics \cite{rotton1985}, in their "meta-analysis of lunar lunacy" produced 37 studies examining correlations between type of lunar cycle, geographical features (latitude, population density, etc.), mental hospital admissions, psychiatric disturbances, crisis calls, homicides, and other criminal offenses. This meta-analysis revealed that alleged associations between moon phases and behavior in previous studies were traced to inappropriate analyses.


This article is organized as follows: \nameref{sec:intro} presents the state of the art in regard to this particular problem, \nameref{sec:methods} described the methods used to assess the association, \nameref{sec:results} describes the results found, \nameref{sec:conclusions} offers an interpretation of the results, and  \nameref{sec:supplemental} presents additional computational support to the claims made in the paper. 

\section*{Methods}\label{sec:methods}
Given the lack of past evidence of any influence of the Moon on human reproduction, we inferred that any effect, if it existed, would be small. Thus, we selected a data set 
whose sample size was big enough to detect very small variations. Human data for this study was obtained from public sources via the US Centers for Disease Control and Prevention (CDC) \cite{CDC}. The CDC collects Vital Statistics for all reported births in the US since 1969. A record is created for each newborn with information including place of birth, residency, parents' socio-economic factors, gestation length, race, birth order, and weight. Starting in 1989, the day of birth was replaced by the week of birth, to de-identify the data. 

The exact date of birth was published for all newborns in the US between 1969 and 1988 (longer than one metonic cycle); 56,545,326 births were recorded in this period. 
We choose a conservative 99\% confidence interval and 90\% power, thus $Z_{\alpha/2}=2.58$ and $Z_{1-\beta} = 1.282$. Our $\mu_0$ and $\mu_1$ differ by $0.0176$ when comparing our two peaks and they each have an approximate standard deviation $\sigma = 0.0429$. Therefore, we estimate that we need to be comparing at least
 $$
   		ES = \frac{\mid \mu_1 - \mu_0 \mid}{\sigma}\quad , \quad
   		n = 2\left( \frac{Z_{\alpha/2}+Z_{1-\beta}}{ES} \right)^2       		= \quad 177.23 \mbox{ \. lunar months.}
 $$
Each year's data is contained in a single flat file in which each record is a continuous chain of characters; the CDC provides documentation to describe the positions and codes of each of the 99 variables. We parsed these 20 years of data (11.3 GB, source code in C++, Transact-SQL, and MATLAB is available as supplemental material) which outputs a comma-separated values (CSV) file containing six additional columns: 
formatted date of birth and conception, the lunar month in which each of these two occurs, and the number of days between a given date and its most recent full moon (computed for conception as well as birth) which we call from now on ``\textit{post-full}'' values. The dates of each full moon were obtained from NASA \cite{NASA}. Between full moons in the synodic cycle, there are 29.53 days on average; thus, we rounded the number of days by considering the last day of the lunar month to contain all births in the interval [28,29.53]. The post-full values range from 0 (when conception or birth occurred on the day of the full moon) up to 28.

	After importing the data into a relational database management system (RDBMS; SQL Server 2012 was used), a number of records per year were randomly selected for comparison against the flat files originated in CDC with no inconsistencies; this quality control procedure gave us confidence in the accuracy of our data transformation. All years of data were stored in a single table in the RDBMS. The relational database file had a size of 69.9 GB. In the RDBMS, we wrote stored procedures that marked the incorrectly reported birth dates with an error code, added a sequential identification column to each record, and altered the tables by adding 11 SQL indexes to speed up queries. Lastly, data was retrieved for analysis with direct calls to the RDBMS from MATLAB. 
    
	It is possible to approximate the date of conception knowing the date of birth, weeks of gestation, birth order, and the age of the mother taking into consideration that: 	
	\begin{itemize}
		\item The number of days of conception varies with birth order; on average full-term pregnancies have 270 days of gestation for primiparous mothers, and 268 days for multiparous mothers\cite{Smith01072001}.
		\item The duration of the menstrual cycle is a function of age\cite{treloar1967} (Figure \ref{fig:treloar}). 
		\item The distribution of the fertile window varies with the duration of the menstrual cycle\cite{Wilcox2000} (Figure \ref{fig:wilcox}). 
	\end{itemize}
	Therefore, to determine the date  of conception, we first recovered data from Treloar \textit{et al.} (1967) and Wilcox \textit{et al.} (2000) using DataThief \cite{datathief} (cf. Figures \ref{fig:treloar} and \ref{fig:wilcox} in Supplemental Material). Then, we used the mother's age to find her likely menstrual cycle length via the median in Treloar \textit{et al.} (1967), and with this information we determined the day of median fertility in Wilcox \textit{et al.} (2000); we call this adjustment in days $A$. This estimation, of course, yields a distribution. Since we have millions of records, we operate under the premise that the central tendency of this distribution reveals the likely true value. 
	
	The date of conception was determined as 
    \[
    	C = B - (7W - A) +\alpha,
    \]
	where $C$ represents the conception date, $B$ represents birth date, $W$ represents the number of weeks of gestation, and $\alpha = 2$ for multiparous mothers or  $\alpha = 0$ otherwise \cite{Smith01072001}. 

	Figure \ref{fig:birthweight} and \ref{fig:gestationRatio} indicate that we must control for gestation. The period between weeks 37 and 41 is considered the time of ``\textit{term births}'', that is, the period considered normal, with 37 weeks general regarded as ``\textit{early term}'', and week 41 usually called ``\textit{late term}''. We constrained our analysis to term births, since this is the period in which it is most likely that births have not been influenced by abnormal factors. 
    
	
	The code retrieved the count of male conceptions on each day and stored them in a 260 x 29 matrix where each column is the count of male conceptions for a particular post-full value and each row is the count of male conceptions from a specific lunar month. The same was done for female conceptions, and then we point-wise divided them to determine the sex ratio of each day. These male:female ratios were stored in a 260 x 29 matrix where each column is ratios for a particular post-full value and each row is all the ratios from a specific lunar month. Each post-full value has a vector of roughly 260 values (a column in the ratio matrix). Figure \ref{fig:stemOverall}(a) displays the average of each of these column vectors. Next, we took the 29 column vectors and ran \textit{z}-tests comparing all 841 possible pairs (cf. MATLAB's pcolor rendering in Figure \ref{fig:conmatrix}). We used a Benjamini-–Hochberg-–Yekutieli procedure with a conservative false discovery rate of 0.01. 
    We also ran the 841 \textit{z}-tests for the male and female matrices. This analysis was repeated for births (Figure \ref{fig:birthmatrix}).
    
	In an effort to further analyze our data for a possible effect of light on sex-determination, we used location data to give our data more filtering capabilities. We were unable to find publicly available data about moonlight. 
    However, cloud cover over North America on average is 3.2\% greater in the daytime than at night, with seasonal variations from 2.0 to 4.3 \cite{hahn1995}. Using sun radiation as an indicator for moonlight, we considered the annual Global Horizontal Irradiance (GHI) average for each county\cite{NREL} (Figure \ref{fig:maps} (a)) and joined this data with the natality data--pairing each birth's NCHS county code of residence with the equivalent FIPS code and GHI annual average in that county. Writing and executing stored procedures in our RDBMS, we determined for each county the number of births reported, the percentage of births for each race, the average sex ratio, average birth weight, average gestation, average mother's age, average father's age, average education of the mother, average education of the father, average start of prenatal care, and average birth order. We used a linear mixed model to further analyze the effects of these variables that displayed slightest correlation.

The linear mixed model is used to investigate the influence of moonlight on sex ratio of the newborn babies. The mixed model refers to the linear model with fixed effects and random effects, which can be used to study the correlated data by grouping the subjects. The fixed effects are of primary interest, which include gestation length and sun light radiation; random effects are thought of as a random selection from groups, which include the groups of population and race.  The fitted linear mixed model is
\begin{equation}
\text{R}_{ijk} = \beta_0 
				+ \beta_1 {G}_{ijk} 
                + \beta_2 {D}_{ijk} 
                + \beta_{12} {(G \cap D)}_{ijk}
                +\gamma_{j}\mathbf{1}
                + \delta_{k} \mathbf{1}
                + \epsilon_{ijk},
\end{equation}
where $G$ represents the period of gestation, $D$ represents radiation, $G \cap D$ represetns the interaction between gestation and radiation; $j$ is the index for population, with $j =1,2,$ and $3$ representing low, medium, and high population density counties, respectively; $i$ is the index for race, with $i =1,2$, and $3$ representing white, black, and Native American, respectively (Hispanics were not consistently reported during the period of study).
$\text{R}_{ijk}$ is the response sex ratio of $i$-th record of population group $j$ and race group $k$; $\beta$'s are the fixed effect parameters, $\gamma$'s are the parameters of the random effect for population groups; $\delta$'s are the parameters of the random effect for groups of race, and $\epsilon_{ijk}$ are the errors.

The fitted values for the parameters of interests are $\hat{\beta_2} = -1.48161$ and $\hat{\beta_{12}} = 0.03751$, which suggest a negative influence of radiation on the sex ratio. To verify the statistical significance of the parameters of interest, $\hat{\beta_2}$ and $\hat{\beta_{12}}$, we performed an ANOVA-type likelihood ratio test, \textit{i.e.} we compare the model with and without radiation effect. The result of this likelihood ratio test is significant with a $p$-value of $0.007107$, which suggests that radiation significantly influences the sex ratio in a negative way. 

We further accounted for artificial light. The majority of artificial light occurs in denser areas. Therefore, using number of births in a county as an indicator of population in that county, we separated our data into three population groups. We sorted the counties by number of births after 1984 (the earlier years had fewer regions reporting) and systematically drew two demarcation lines (which were at 2834 and 3082 births per county) such that each of the three groups determined by these divisions comprises approximately one-third of the population---these sets contain different quantities of counties. Group 1 contains the smallest counties; the medium size counties comprise group 2; group 3 is the set of the largest counties. We added a table in our RDBMS that contains each county code and its population group assignment. This population group filter was added to our analysis, and we used it to compare high-density population areas to low-density areas, intending to compare urban areas with high artificial light to rural areas with little artificial light.

Lastly, we added the filtering capability to choose random birth entries to objectively choose a random subset for analysis. We wrote a stored procedure in our RDBMS that randomly assigns floating point values to each of our birth records. These random numbers follow a Gaussian distribution. After this assignment, every one of the 56 million birth records has a random value by which it will be filtered. For example, when we filter by 1 standard deviation, then every birth record which has a Gaussian value within 1 standard deviation of the mean will be included in the set which we then run analysis on. We repeated this filtering and analysis additionally on all records with a value within 2 standard deviations of the mean.




\section*{Results}\label{sec:results}

\begin{figure}
	\centering
		\includegraphics[width=6in]{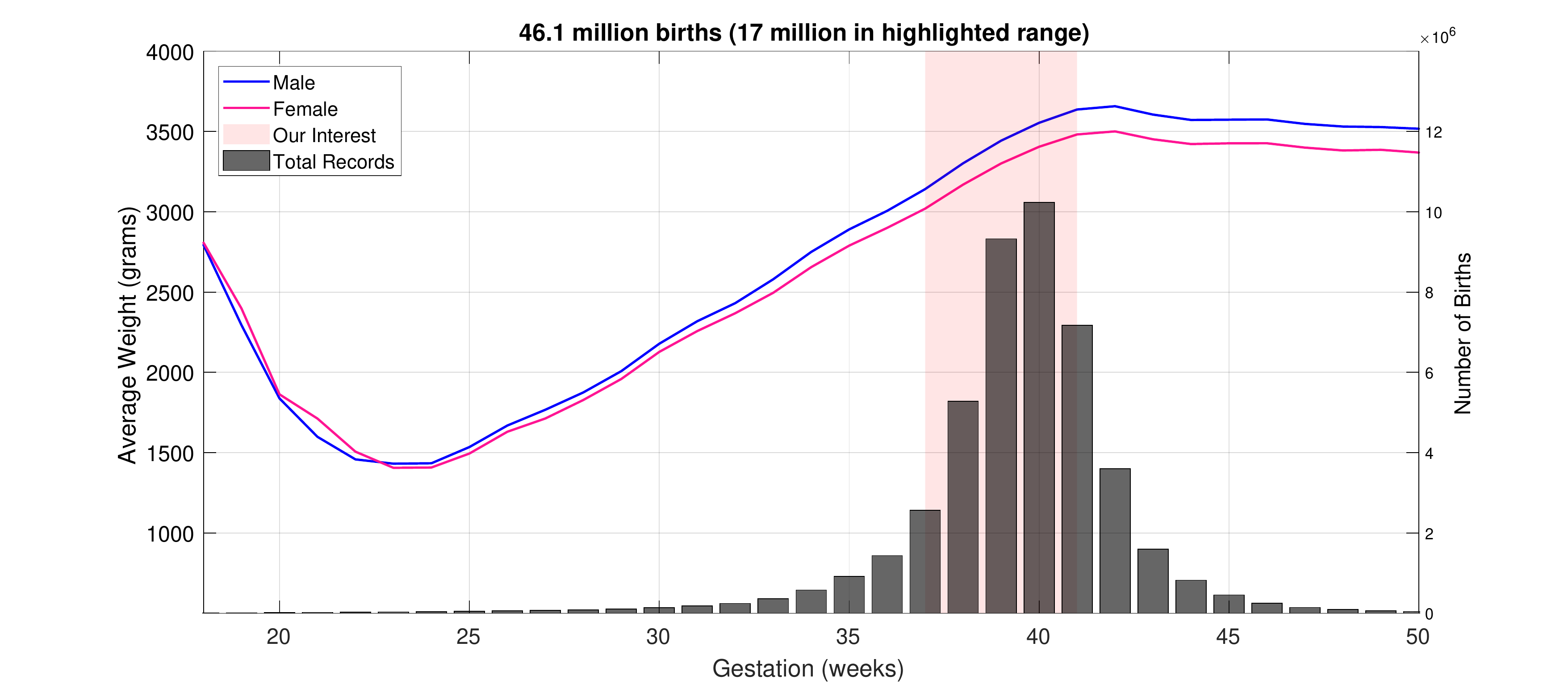}
	\caption{Average birth weight for each sex per gestation length. Additionally, the number of total records for each gestation length are shown, and our filtered window of 30 to 40 weeks is displayed.}
	\label{fig:birthweight}
\end{figure}

\begin{figure}
	\centering
		\includegraphics[width=6in]{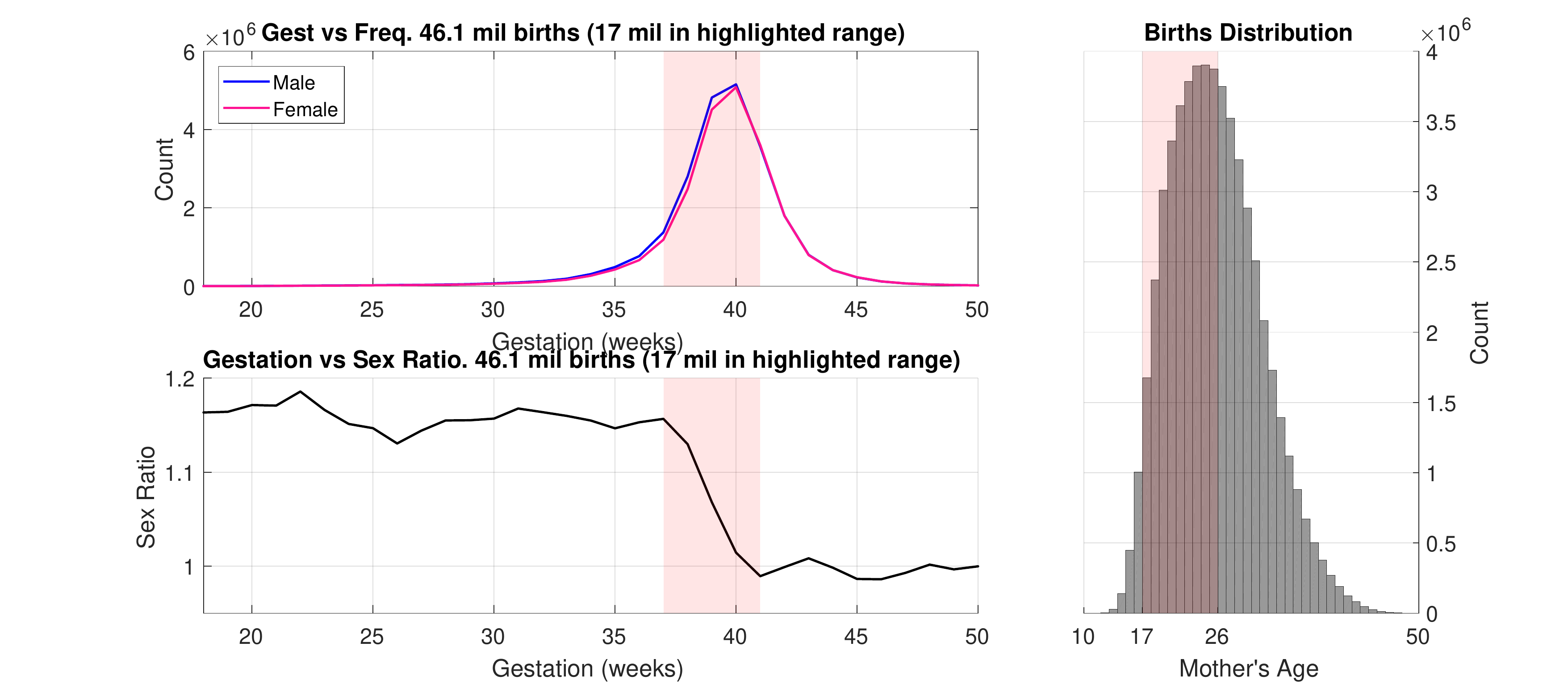}
	\caption{Frequency and sex ratio varying with gestation. (i) Frequency of births for each gestation length and sex. (ii) Sex ratio for each gestation length. Our filtered window of 30 to 40 weeks is shown in  both.}
	\label{fig:gestationRatio}
\end{figure}

\begin{figure}
	\centering
		\includegraphics[width=6in]{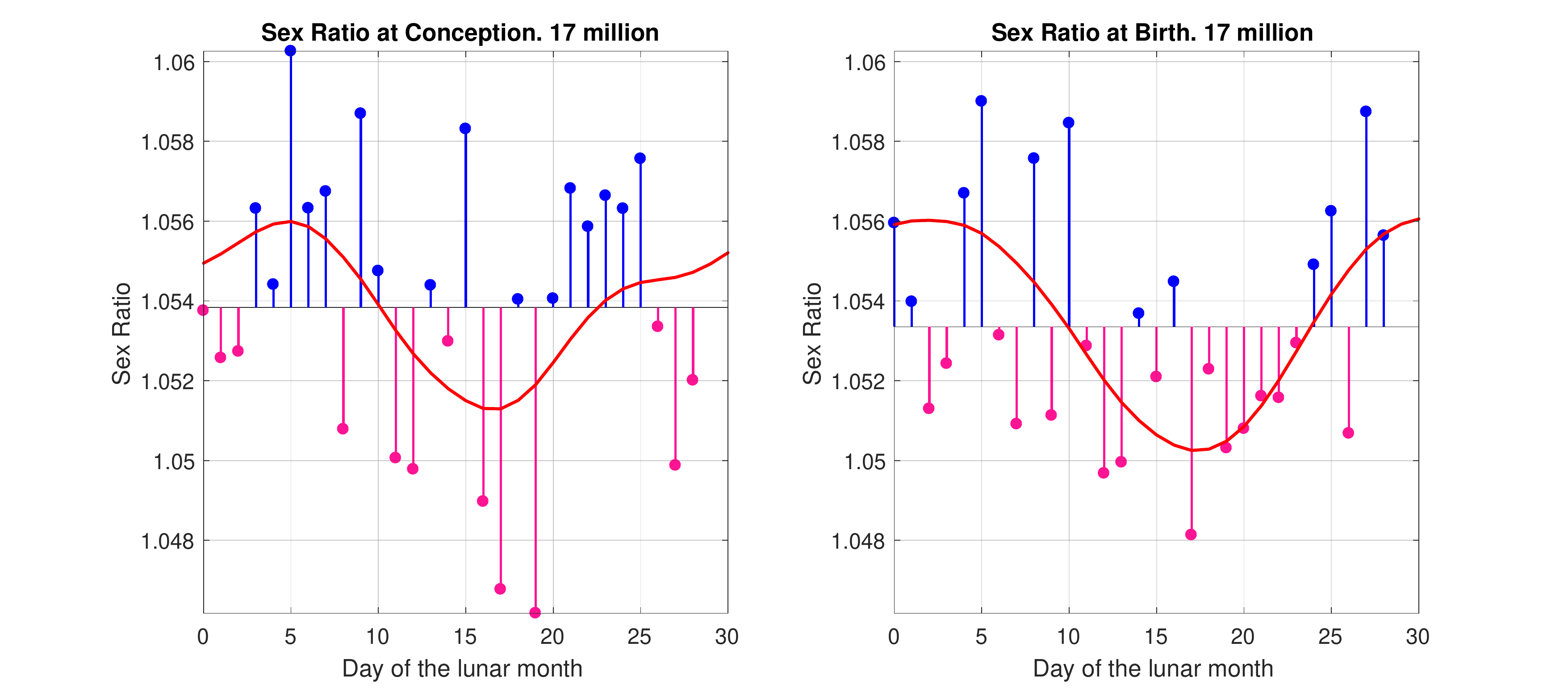}
	\caption{Sex ratio on each day of the lunar month for (i) conception and (ii) birth, where gestation is between 30 and 40 weeks. Days of the lunar month that favor boys more than the mean are blue. Days of the lunar month that favor girls more than the mean are pink. }
	\label{fig:stemOverall}
\end{figure}

\begin{figure}
	\centering
		\includegraphics[width=6in]{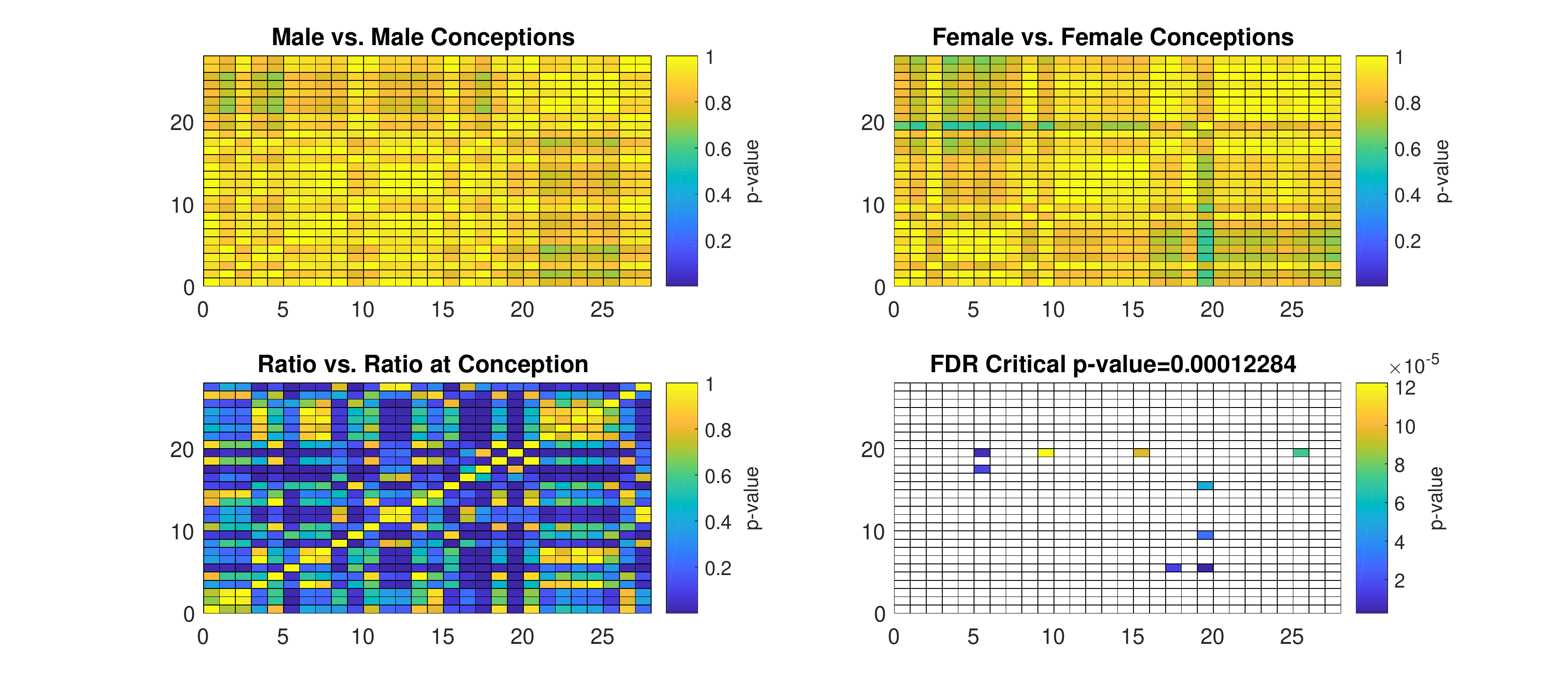}
	\caption{Comparison matrices of \textit{z}-test \textit{p}-values for all pairs of post-full conception values.  Number of male conceptions, number of female conceptions, and sex ratios were compared. The fourth matrix colors just the values of the ratio comparison matrix lower than the critical \textit{p}-value. The coloring of the small squares represents an interpolated value of its corners.}
	\label{fig:conmatrix}
\end{figure}
    
\begin{figure}
	\centering
		\includegraphics[width=6in]{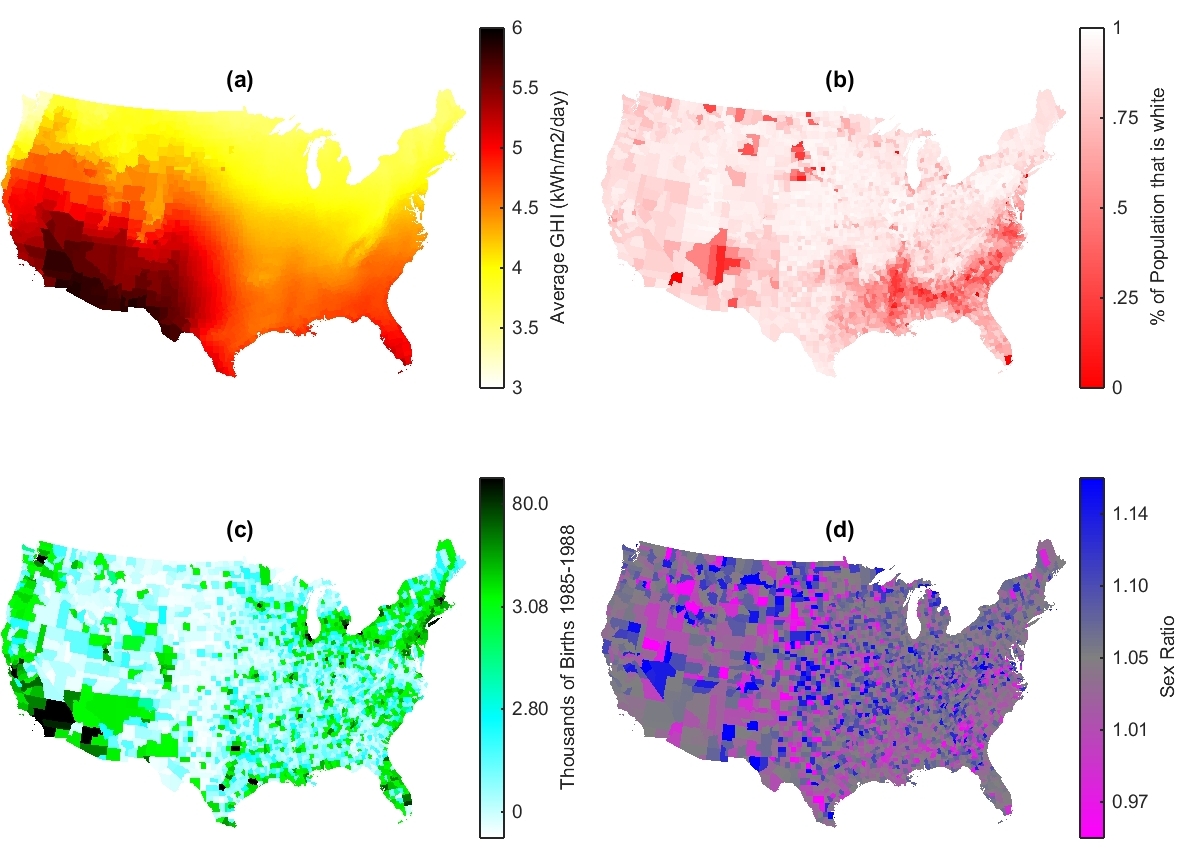}
	\caption{(a) GHI by County (b) percentage of the babies born to two white parents as an indicator of white percentage of the population (c) Total number of births in each county as an indicator for county population with population groups [0,2834],(2834,3082],(3058,659519] (d) Sex ratio for each county, where sex ratio mean is 1.0534 and standard deviation is 0.0421}
	\label{fig:maps}
\end{figure}



	We determined a gestation length restriction for our analysis by observing how birth weight and sex ratio vary with gestation. We graphed the average birth weight for each sex at each length of gestation (in integer weeks) (Figure \ref{fig:birthweight}). On this graph, the number of total records for each gestation length is shown in bar form and our interest window of 37-41 weeks is shaded. We next showed the relationship between gestation length and total births for each sex (Figure \ref{fig:gestationRatio}(i)); the same was done for gestation and sex ratio immediately below (Figure \ref{fig:gestationRatio}(ii)).
    
    For every day of the lunar month, the average sex ratio was calculated and plotted for the conceptions occurring on that day; this was repeated for births for contrast (Figure \ref{fig:stemOverall}). When the average sex ratio for a given day of the lunar month has a value greater than the mean of all days, it favors males more than the mean and is colored blue. If a day has an average sex ratio less than the total mean, it is colored pink. Although we use the same data and records, the means of the sex ratios by lunar day are not equal because we are calculating the mean of the daily mean of ratios, which are non-trivially altered by the conception date calculation. More specifically, the sex ratio at birth on a given day does not have a precisely corresponding value in our conception matrix roughly 9 months prior; thus, averaging different ratios gives us two different means. (As a brief example, suppose on each day you have [number of male, number of female] births for two days.  Day 1 is [3,3], and Day 2 is [3,2]. You then calculate the conception dates A and B with the slightest variation for these 11 babies: Day A [2,4] , and Day B is [4,1]. Our sex ratios per day at birth are 3/3 and 3/2 with an average of the ratios as 1.25. At conception, our sex ratios are 2/4 and 4/1 with an average of these ratios as 2.25. Notice that the overall sex ratio is 6 males to 5 females which results in a 1.2 sex ratio. All three of these are different.)
    
    We ran \textit{z}-tests comparing all possible pairs of post-full series values and plotted the resulting \textit{p}-values (Figure \ref{fig:conmatrix}). We compared male counts, female counts, and sex ratios. The fourth matrix colors just the $p$-values of the ratio comparison matrix lower than the critical \textit{p}-value. Every small square represents an interpolated value of its four vertices and is colored as such.
    
    We further explored other filters for our analysis to understand the correlations we observed. We displayed the sun radiation, specifically the GHI average, for each county. Because we wish to consider other factors, we also mapped the percentage of the babies born to white parents as an indicator for the percentage of the population that is white, the total number of births in each county as an indicator for population, and the average sex ratio for each county (Figure \ref{fig:maps}).
    

\section*{Discussion}\label{sec:conclusions}

We witness a distinctive variation in birth weight, sex ratio, and quantity of births for each gestation length.  
we observed  that for a given gestation length males weighed more than females at birth, on average (Figure \ref{fig:birthweight}). Additionally, prior to week 37, some unidentified phenomenon favors the birth of approximately 15\% more males than females in a very consistent way. This phenomenon stops in week 37, and then we observe a fast decrease of sex ratios, so that by week 42, the same number of males and females are born (Figure \ref{fig:gestationRatio}). 
It is interesting to note that the average birth weights at this same time cease increasing (Figure \ref{fig:birthweight}), which gives weight to the concept that the incredibly male-dominant sex ratio for pre-term births stems from males growing faster than females. Once growth stops around 40 weeks, male and female births even out.

However, more male than female births in an aggregate 46 million records is not as convincing as a daily analysis. Thus, we observed the frequency (using a Hodrick-Prescott filter) of male and female reported births on every single day between 1969 and 1988 as well as the count of male and female computed conceptions in the window roughly nine months prior (see Supplemental Material Figure \ref{fig:dailyfrequencies}). Males outnumber females at a relatively regular rate. 
Additionally, Americans appear to follow a seasonal reproductive cycle with peak conception in January and June (peak births in March and October).
  
Noticeable spikes (the greatest variations from the mean sex ratio) exist for conceptions 5 days and 19 days after a full moon. Conception of a male is most probable five days after a full moon (Figure \ref{fig:stemOverall}). The best chance for conceiving a female is 19 days after a full moon, although the sex ratio still favors males at this time. 
Notice that the same is not true for births. In fact, the graph of births does not display a similar or shifted representation whatsoever. Our data agrees with the observation that noticeably more male births occur on full moon than female births.

We need to understand if the spikes we are seeing are significant. We witness an incredibly significant \textit{p}-value of $8.0934 \times 10^{-11}$ for the \textit{z}-test comparing day 5 against day 19 (and $1.1\times 10^{-10}$ for 19 against day 5) (Figure \ref{fig:birthmatrix}). The same \textit{z}-tests were run for the births (Figure \ref{fig:birthmatrix} in Supplemental Material). The different results indicate that the conception significance is not an artifact originating from the conception calculation. In fact, the peak at day 27 is significant; the \textit{z}-test comparing day 27 to day 19 has a \textit{p}-value of $6.2 \times 10^{-7}$.

Our readings about the pineal gland suggest that moonlight might play a role in this correlation. Operating under the assumption that cloud cover at night follows the same trends as the day, our intent is to use sun radiation measurements to determine the relative intensities of moonlight in different areas. We plotted data from the National Renewable Energy Laboratory (NREL) \cite{NREL} which presented the Global Horizontal Irradiance (GHI) average for each county (Figure \ref{fig:maps} (a)). This solar radiation data implements the SUNY model, which accounts for atmospheric water vapor, trace gases, and aerosols (nontrivial in large cities). This model calculates similar GHI values as the MAC3 model which accounts for cloud cover. 
For additional visualization, we mapped the percentage of the white population per county (Figure \ref{fig:maps}(b)) to visualize that our observation is not confounded by the race demographics of regions in the US. 
We also mapped the population per county (with color selection to follow our population group assignment) (Figure \ref{fig:maps}(c)) so we could see where the artificial light may be affecting our observations. Nonetheless, although the average total sex ratio per county has a visual signature that appears uncorrelated to region, our linear mixed model indicates otherwise (Figure \ref{fig:maps}(d)).

To further visualize the proposed effect of moonlight on sex ratio, we filtered the data and ran the analysis on these subgroups, ensuring that each subgroup exceeded our sample size requirement. 
In an effort to compare rural counties to urban counties (which tend to possess more light pollution), we split the data into three groups, each containing roughly one-third of the births in the 20 years period. We use number of births to residents of a county as an indicator for population. 
Notice that in the counties with the highest GHI radiation, the effect of the lunar cycle on sex ratio is more pronounced. However, for most of these subgroups, few significant correlations can be made. 
The highly populated counties are the exception. Numerous significant \textit{p}-values display a similar pattern of the overall and white overall analyses. The abundance of light at night in urban areas, could be suppressing melatonin which gives us this more pronounced signature. We repeated this analysis for births (Figure \ref{fig:stemBirth} in Supplemental Material), which shows once again that we are not observing an artifact based on our conception calculation and that in all subgroups, births right before the full moon noticeably favor males.

In conclusion, the lunar cycles seem to have a very significant influence on sex determination in humans for a specific segment of mothers (ages 17-26, full-term pregnancies).

\section*{Supplemental Material}\label{sec:supplemental}
\setcounter{figure}{0} 
\renewcommand{\thefigure}{S\arabic{figure}}

\begin{figure}
	\centering
		\includegraphics[width=6in]{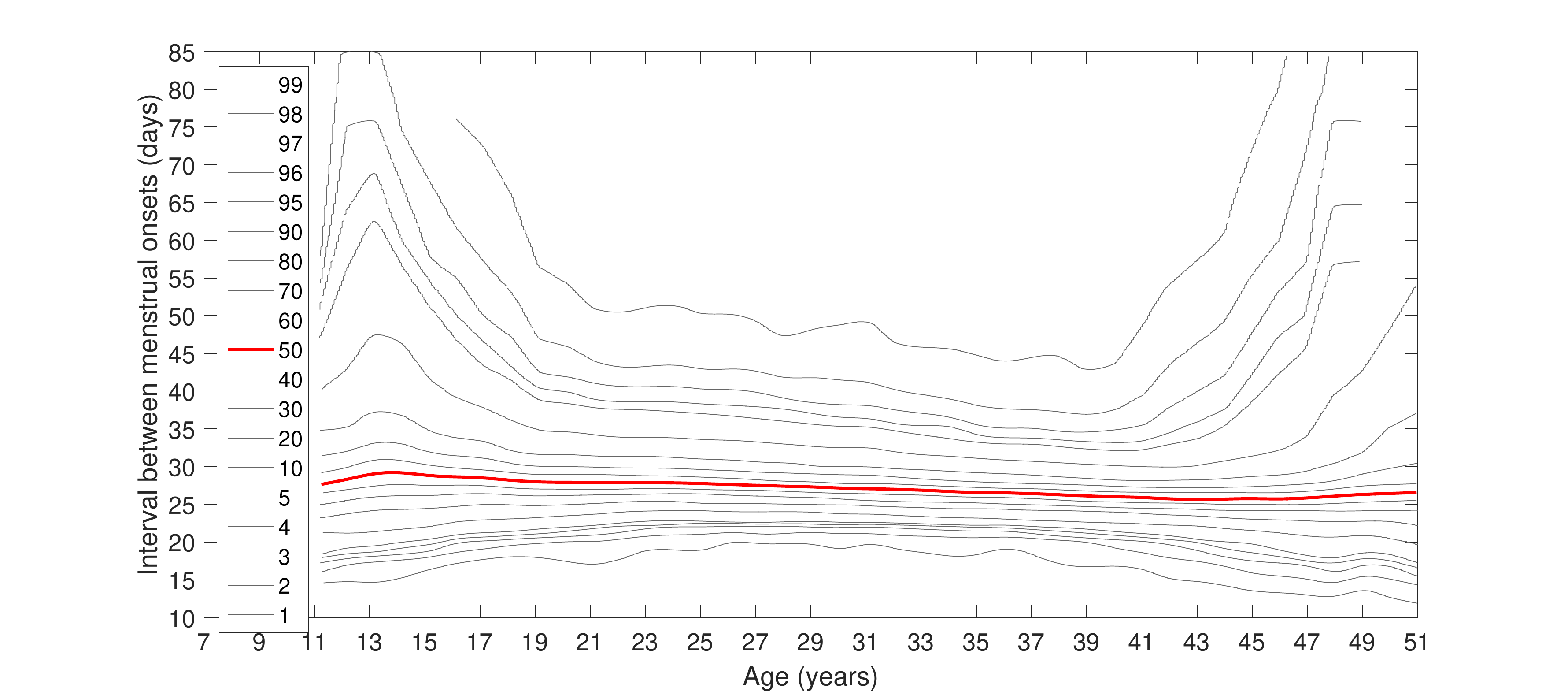}
	\caption{A representation showing the variation in the menstrual cycle length with the age of the female. The median of this graph is highlighted in red. The other percentile groups are shown in gray. This graph is a replica of the data yielded by Treloar (1967) \cite{treloar1967}.}
	\label{fig:treloar}
\end{figure}

\begin{figure}
	\centering
		\includegraphics[width=6in]{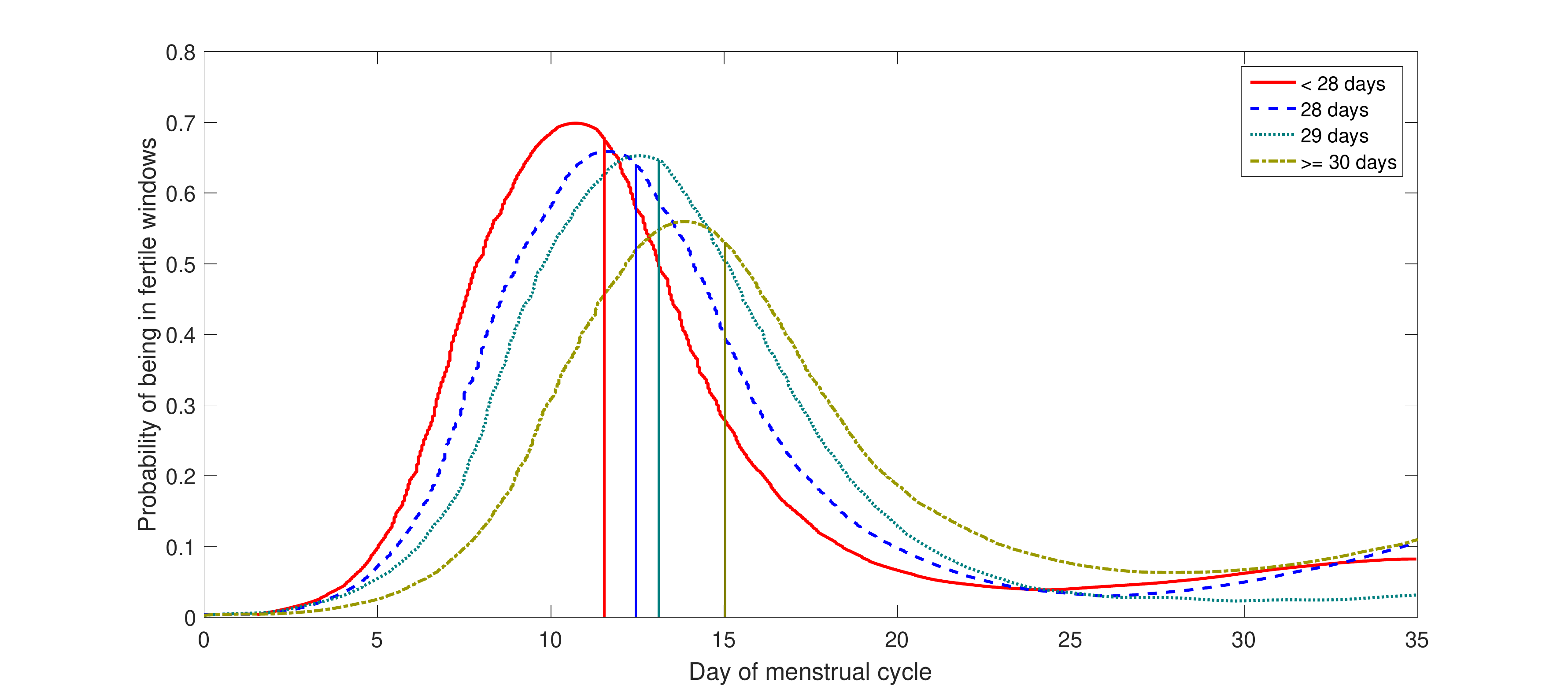}
	\caption{Peak fertility curves during the menstrual cycle for females of different menstrual lengths taken from Wilcox (2000)\cite{Wilcox2000}. The median of each curve is displayed and is the most likely day of the menstrual cycle on which conception occurs.}
	\label{fig:wilcox}
\end{figure}

\begin{figure}
	\centering
		\includegraphics[width=6in]{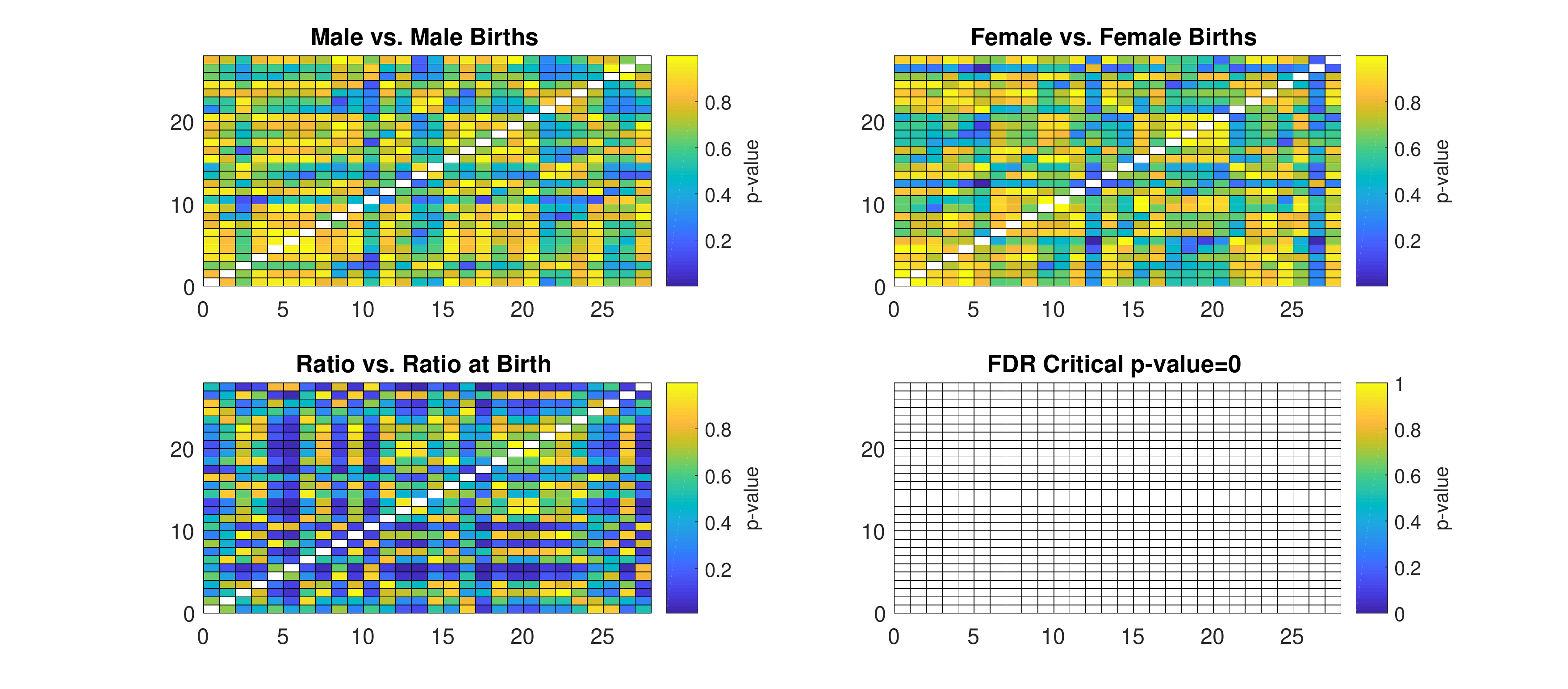}
	\caption{Comparison matrices of \textit{z}-test \textit{p}-values for all pairs of post-full birth values. Number of male births, number of female births, and sex ratios were compared. The fourth matrix shows just the values of the ratio comparison matrix lower than the critical \textit{p}-value. The coloring of the small squares represents an interpolated value of its corners.}
	\label{fig:birthmatrix}
\end{figure}

\begin{figure}
	\centering
		\includegraphics[width=6in]{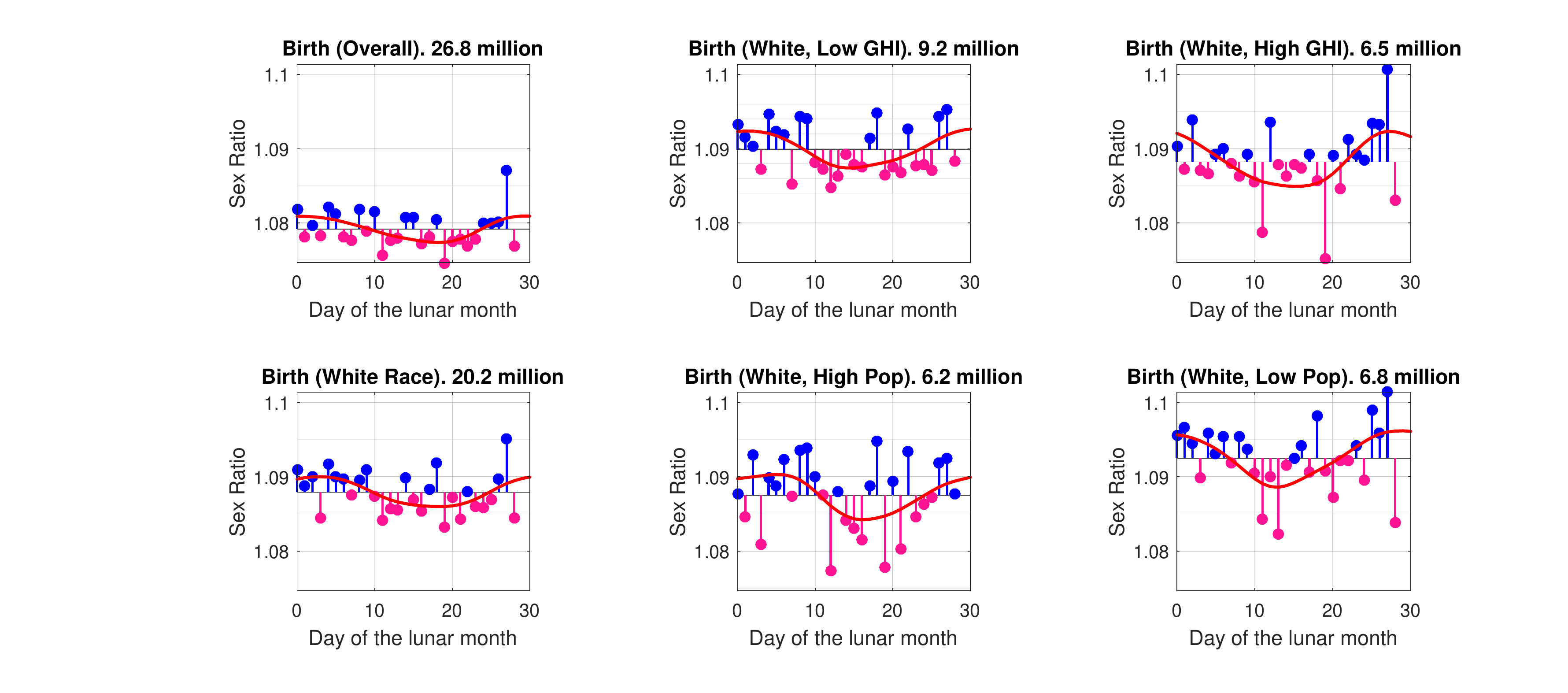}
	\caption{Sex ratios at birth per lunar day each using a different filter in our interest period. We display the overall stem plot from \ref{fig:stemOverall} for comparison, a stem plot filtered only for babies born to two white parents, and four other plots that contain the race filter in addition to other filters: low GHI, high GHI, low population, and high population. counties separated by GHI are put into three groups based on GHI levels. The lowest third of counties comprise the low group, and the highest third comprise the high group. Using number of births as an indicator for population, the low population group contains the counties with the fewest births that total one-third of total births. The high population group contains the counties with the most births that total one-third of total births. }
	\label{fig:stemBirth}
\end{figure}

\begin{figure}
	\centering
		\includegraphics[width=6in]{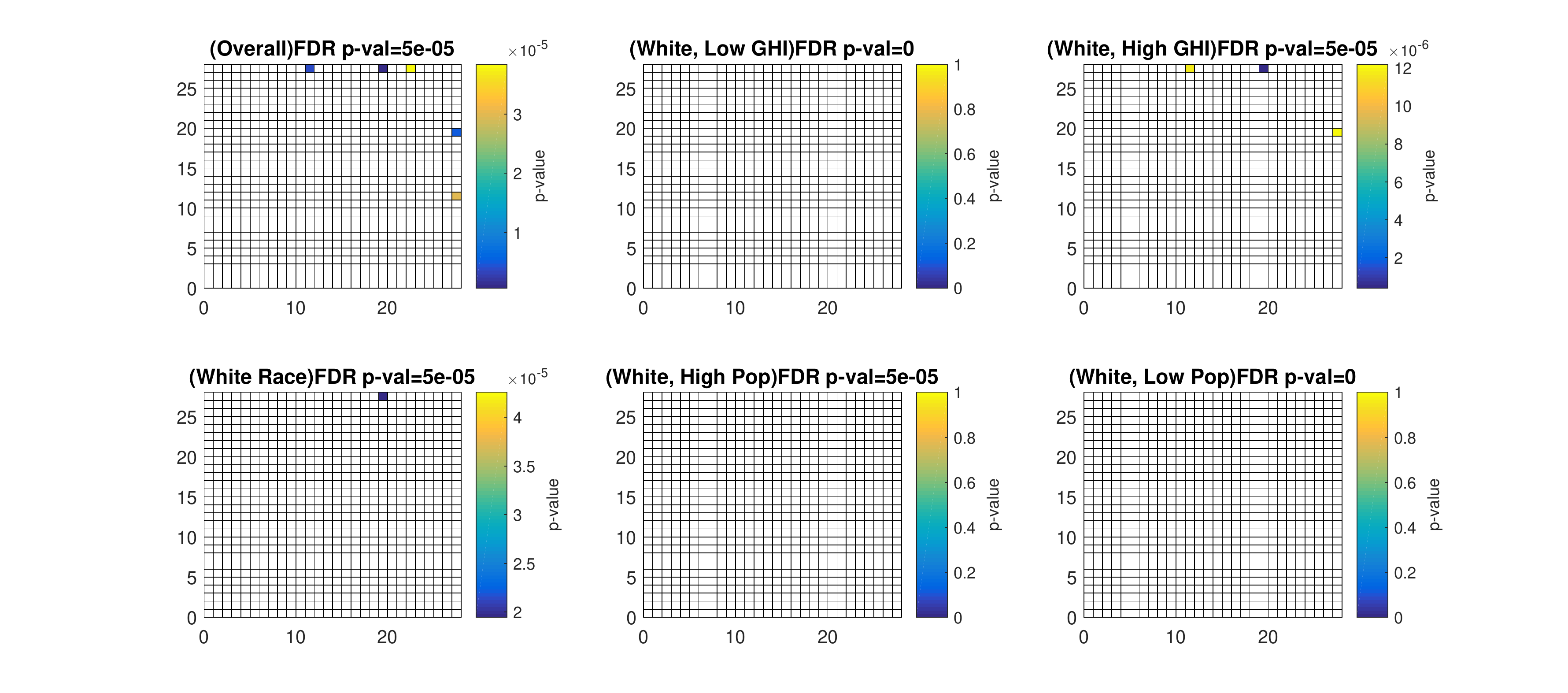}
	\caption{The matrices of \textit{p}-values of the \textit{z}-tests for the plots in Figure \ref{fig:stemBirth} using the methodology of the final matrix in Figure \ref{fig:conmatrix}.}
	\label{fig:BirthMatrices}
\end{figure}

\begin{figure}
	\centering
		\includegraphics[width=6in]{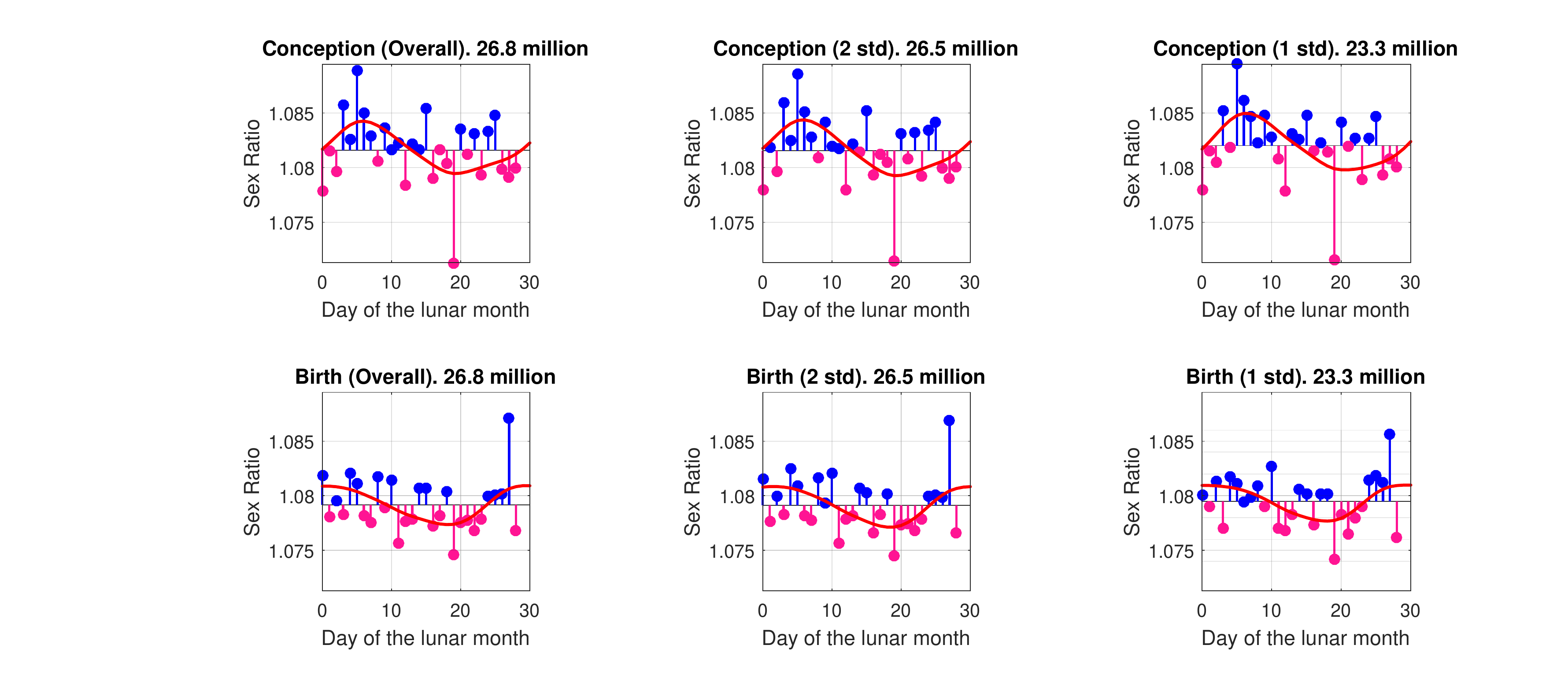}
	\caption{Sex ratios at conception and birth per lunar day using the Gaussian random number filter in our interest period. All records with a random value within 2 standard deviations from the mean and within 1 standard deviation are displayed. The overall stem plots from our interest period are included for comparison.}
	\label{fig:gaussianStems}
\end{figure}

\begin{figure}
	\centering
		\includegraphics[width=6in]{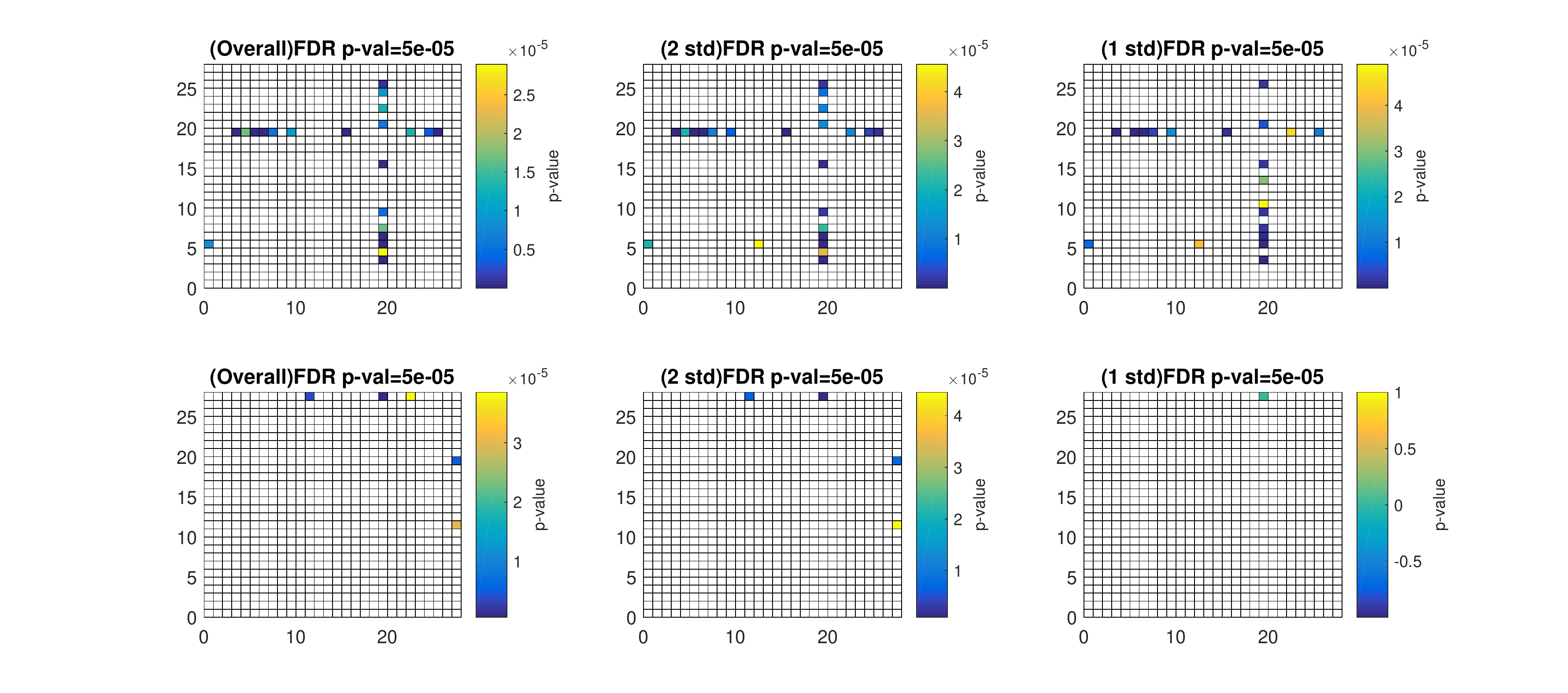}
	\caption{The matrices of \textit{p}-values of the \textit{z}-tests for the plots in Figure \ref{fig:gaussianStems} using the methodology of the final matrix in Figure \ref{fig:conmatrix}.}
	\label{fig:gaussianMatrices}
\end{figure}

\begin{figure}
	\centering
		\includegraphics[width=6in]{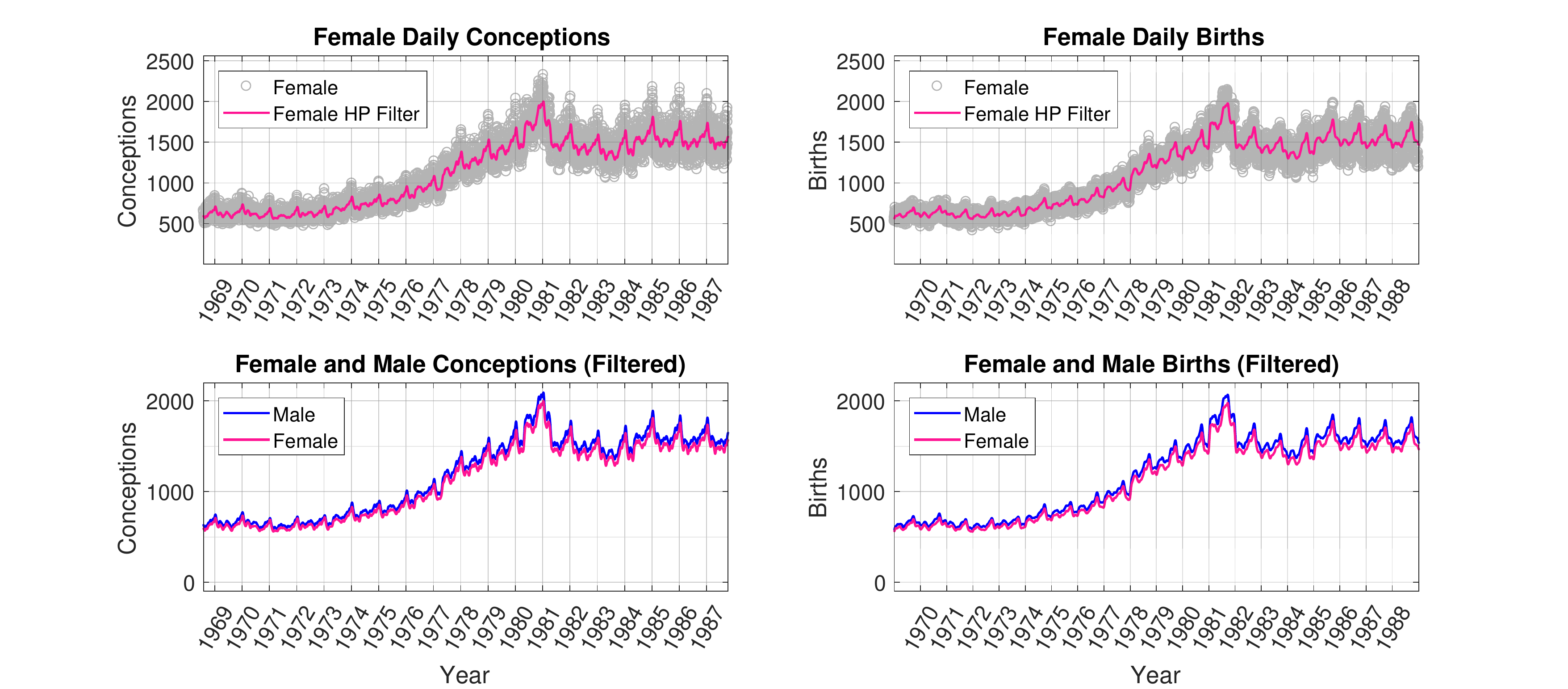}
	\caption{Frequency of births per day}
	\label{fig:dailyfrequencies}
\end{figure}

Many authors have looked into the seasonality of births and conceptions. Although we discussed the seasonality of American births, Zimecki's study shows that the French deliver the greatest number of births in May and the least around September and October \cite{zimecki2006}. Roenneburg and Aschoff (1990) \cite{roenneburg1990} state that a peak of conceptions occurs near the spring equinox as an aggregate result of their 166 geographical regions with birth data from the past 3000 years. Their 9 month difference calculation indicates this from their peak of births in December. Their specific data for the District of Columbia from 1908-1909 displays peaks in births in February and July which indicated peak conceptions in May and October. Our findings (Figure \ref{fig:dailyfrequencies}), however, indicate peaks in American conceptions occurring in January and June and peaks in deliveries occurring in March and October. 

Here, we analyzed the effect of moonlight on sex ratio. We did not yet look into gravity.
The X chromosome is 3\% heavier than the  Y chromosome; we hypothesize that this is another possible reason for the significant peaks we are witnessing. Analysis taking into account the gravitational forces of the three-body problem \cite{gutzwiller1998} and the anomalistic cycle needs to be done to determine if gravity is a factor.


Our analysis of this dataset is by no means complete.
Cutler \textit{ et al.} (1987) reported a relationship between the onset of menstruation and the onset of full moon; if prevalent, more conceptions would occur during new moon. Our rough findings indicated no significant correlation between total conceptions and lunar phase. 
Furthermore, we saw no peaks in total births in the few days following the full moon, where Ghiandoni \textit{et al.} (1998) \cite{ghiandoni1998} claims peaks occur. Meanwhile,  Zimecki (2006) \cite{zimecki2006} discusses births varying with week day. Our non-rigorous look into this revealed non-significant variations. These analyses have much better framing in a more rigorous and separate paper. 


\begin{thebibliography}{10}

\bibitem{NREL}
{National Renewable Energy Laboratory, a national laboratory of the U.S.
  Department of Energy}.
\newblock \url{http://www.nrel.gov/gis/data_solar.html}.
\newblock Accessed: July 14, 2016.

\bibitem{NASA}
{Sky Events Calendar by Fred Espenak and Sumit Dutta (NASA's GSFC)}.
\newblock \url{http://eclipse.gsfc.nasa.gov/SKYCAL/SKYCAL.html}.
\newblock Accessed: January, 2014.

\bibitem{CDC}
{US Center for Disease Control and Prevention, National Vital Statistics
  System}.
\newblock \url{http://www.cdc.gov/nchs/nvss.htm}.
\newblock Accessed: October 9, 2013.

\bibitem{charpentier2009}
A.~Charpentier and D.~Causeur.
\newblock {Large-scale significance testing of the full Moon effect on
  deliveries}.
\newblock working paper or preprint, Mar. 2009.

\bibitem{cochran1988}
R.~Cochran, S.~Zabludoff, K.~Paynter, L.~DiMichele, and R.~Palmer.
\newblock Serum hormone levels associated with spawning activity in the
  mummichog, fundulus heteroclitus.
\newblock {\em General and comparative endocrinology}, 70(2):345--354, 1988.

\bibitem{cutler1987}
W.~B. Cutler, W.~M. Schleidt, E.~Friedmann, G.~Preti, and R.~Stine.
\newblock {{L}unar influences on the reproductive cycle in women}.
\newblock {\em Hum. Biol.}, 59(6):959--972, Dec 1987.

\bibitem{Das2005}
S.~Das, S.~Dodd, D.~I. Lewis-Jones, F.~M. Patel, A.~J. Drakeley, C.~R.
  Kingsland, and R.~Gazvani.
\newblock Do lunar phases affect conception rates in assisted reproduction?
\newblock {\em Journal of Assisted Reproduction and Genetics}, 22(1):15--18,
  2005.

\bibitem{foster2008}
R.~G. Foster and T.~Roenneberg.
\newblock {{H}uman responses to the geophysical daily, annual and lunar
  cycles}.
\newblock {\em Curr. Biol.}, 18(17):R784--R794, Sep 2008.

\bibitem{fox1924lunar}
H.~M. Fox.
\newblock Lunar periodicity in reproduction.
\newblock {\em Proceedings of the Royal Society of London. Series B, Containing
  Papers of a Biological Character}, 95(671):523--550, 1924.

\bibitem{ghiandoni1998}
G.~Ghiandoni, R.~Secli, M.~B. Rocchi, and G.~Ugolini.
\newblock {{D}oes lunar position influence the time of delivery? {A}
  statistical analysis}.
\newblock {\em Eur. J. Obstet. Gynecol. Reprod. Biol.}, 77(1):47--50, Mar 1998.

\bibitem{grau1981}
E.~G. Grau, W.~W. Dickhoff, R.~S. Nishioka, H.~A. Bern, and L.~C. Folmar.
\newblock {{L}unar phasing of the thyroxine surge preparatory to seaward
  migration of salmonid fish}.
\newblock {\em Science}, 211(4482):607--609, Feb 1981.

\bibitem{gutzwiller1998}
M.~C. Gutzwiller.
\newblock Moon-earth-sun: The oldest three-body problem.
\newblock {\em Rev. Mod. Phys.}, 70:589--639, Apr 1998.

\bibitem{hahn1995}
C.~J. Hahn, S.~G. Warren, and J.~London.
\newblock The effect of moonlight on observation of cloud cover at night, and
  application to cloud climatology.
\newblock {\em Journal of Climate}, 8(5):1429--1446, 1995.

\bibitem{kollerstrom2000}
N.~Kollerstrom and C.~Power.
\newblock The influence of the lunar cycle on fertility on two thoroughbred
  studfarms.
\newblock {\em Equine Veterinary Journal}, 32(1):75--77, 2000.

\bibitem{littleton2002}
Littleton, editor.
\newblock {\em Mythology: The Illustrated Anthology of World Myth and
  Storytelling}.
\newblock Duncan Baird Publishers, 2002.

\bibitem{martinez2002}
F.~Martinez-Soriano, E.~Armananzas, A.~Ruiz-Torner, and A.~A. Valverde-Navarro.
\newblock {{I}nfluence of light/dark, seasonal and lunar cycles on the nuclear
  size of the pinealocytes of the rat}.
\newblock {\em Histol. Histopathol.}, 17(1):205--212, Jan 2002.

\bibitem{navara2007}
K.~J. Navara and R.~J. Nelson.
\newblock The dark side of light at night: physiological, epidemiological, and
  ecological consequences.
\newblock {\em Journal of Pineal Research}, 43(3):215--224, 2007.

\bibitem{Perea2005}
F.~Perea and E.~Soto.
\newblock Influencia del ciclo lunar sobre la fertilidad en vacas mestizas de
  doble prop\'osito.
\newblock In {\em Memorias 6 Simposio Internacional de Reproducción Animal},
  page 412, 24-26 Junio 2005.

\bibitem{periti1994}
E.~Periti and R.~Biagiotti.
\newblock {[{L}unar phases and incidence of spontaneous deliveries. {O}ur
  experience]}.
\newblock {\em Minerva Ginecol}, 46(7-8):429--433, 1994.

\bibitem{roenneburg1990}
T.~Roenneberg and J.~Aschoff.
\newblock {{A}nnual rhythm of human reproduction: {I}. {B}iology, sociology, or
  both?}
\newblock {\em J. Biol. Rhythms}, 5(3):195--216, 1990.

\bibitem{rotton1985}
J.~Rotton and I.~W. Kelly.
\newblock {{M}uch ado about the full moon: a meta-analysis of lunar-lunacy
  research}.
\newblock {\em Psychol Bull}, 97(2):286--306, Mar 1985.

\bibitem{Smith01072001}
G.~C. Smith.
\newblock Use of time to event analysis to estimate the normal duration of
  human pregnancy.
\newblock {\em Human Reproduction}, 16(7):1497--1500, 2001.

\bibitem{strolego1991}
F.~Strolego, C.~Gigli, and A.~Bugalho.
\newblock {[{T}he influence of lunar phases on the frequency of deliveries]}.
\newblock {\em Minerva Ginecol}, 43(7-8):359--363, 1991.

\bibitem{takemura2004}
A.~Takemura, M.~S. Rahman, S.~Nakamura, Y.~J. Park, and K.~Takano.
\newblock Lunar cycles and reproductive activity in reef fishes with particular
  attention to rabbitfishes.
\newblock {\em Fish and Fisheries}, 5(4):317--328, 2004.

\bibitem{takemura2010}
A.~Takemura, M.~S. Rahman, and Y.~J. Park.
\newblock {{E}xternal and internal controls of lunar-related reproductive
  rhythms in fishes}.
\newblock {\em J. Fish Biol.}, 76(1):7--26, Jan 2010.

\bibitem{treloar1967}
A.~E. Treloar, R.~E. Boynton, B.~G. Behn, and B.~W. Brown.
\newblock {{V}ariation of the human menstrual cycle through reproductive life}.
\newblock {\em Int. J. Fertil.}, 12(1 Pt 2):77--126, 1967.

\bibitem{datathief}
B.~Tummers.
\newblock Datathief iii.
\newblock \url{http://datathief.org/}, 2006.
\newblock Accessed: March 22, 2016.

\bibitem{Wilcox2000}
A.~J. Wilcox, D.~Dunson, and D.~D. Baird.
\newblock The timing of the {\textquotedblleft}fertile
  window{\textquotedblright} in the menstrual cycle: day specific estimates
  from a prospective study.
\newblock {\em BMJ}, 321(7271):1259--1262, 2000.

\bibitem{zimecki2006}
M.~Zimecki.
\newblock {{T}he lunar cycle: effects on human and animal behavior and
  physiology}.
\newblock {\em Postepy Hig Med Dosw (Online)}, 60:1--7, 2006.

\end{thebibliography}


\section*{Acknowledgments}
The authors thank Lisa Jagoda, Department of Mathematics at UGA for her help in translating Aristotle.

\section*{Author contributions statement}
J.B.G. conceived the experiment,  J.A. advised in the statistical analysis, D.O. did the data processing, E.M. advised on the visualization, R.P. assisted in the biological interpretation of the data, and R.X. and L.Z developed the mixed linear model.  All authors reviewed the manuscript. 

\section*{Additional information}
The authors declare no competing financial interests.

\end{document}